\documentclass[5p,times]{elsarticle}
\usepackage{graphicx}
\usepackage{amsmath}
\usepackage[english]{babel}
\usepackage{amsfonts,amssymb}
\usepackage[mathscr]{eucal}

\begin{document}

\title{Twisted magnetization states and inhomogeneous resonance modes\\
in a Fe/Gd ferrimagnetic multilayer}

\author[ipp]{A.B.~Drovosekov\corref{cor}}
\ead{drovosekov@kapitza.ras.ru}
\author[ipp,issp]{A.O.~Savitsky}
\author[ipp]{D.I.~Kholin}
\author[ipp]{N.M.~Kreines}
\author[imp]{V.V.~Proglyado}
\author[imp]{M.V.~Ryabukhina}
\author[imp,ufu]{E.A.~Kravtsov}
\author[imp,ufu]{V.V.~Ustinov}

\cortext[cor]{Corresponding author}

\address[ipp]{P.L.~Kapitza Institute for Physical Problems RAS, 119334 Moscow, Russia}

\address[issp]{Institute of Solid State Physics RAS, 142432 Chernogolovka, Moscow region, Russia}

\address[imp]{M.N.~Mikheev Institute of Metal Physics UB RAS, 620137 Ekaterinburg, Russia}

\address[ufu]{Ural Federal University, 620002 Ekaterinburg, Russia}

\begin{abstract}
Static and dynamic magnetic properties of a ferrimagnetic [Fe(35\AA)/Gd(50\AA)]$_{12}$ superlattice were investigated in a wide $4-300$~K temperature range using magneto-optical Kerr effect (MOKE) and ferromagnetic resonance (FMR) techniques. The multilayer structure was sputtered on a transparent glass substrate which made it possible to perform MOKE measurements on both Fe and Gd terminated sides of the superlattice. These experiments allowed us to detect a transition between field-aligned and canted magnetic states on both sides of the film and to distinguish between the bulk and surface twisted phases of the superlattice. As a result, the experimental $H-T$ magnetic phase diagram of the system was obtained. FMR studies at frequencies $7-36$~GHz demonstrated a complex evolution of absorption spectra as temperature decreased from room down to 4~K. Two spectral branches were detected in the sample. Theoretical simulations show that the observed spectral branches correspond to different types of inhomogeneous resonance modes in the multilayer with non-uniform magnetization precession inside Gd layers.
\end{abstract}


\maketitle

\section{Introduction}

Layered structures based on transition (TM) and rare-earth (RE) ferromagnetic (FM) metals, like Fe/Gd, are model ferrimagnetic systems demonstrating a rich magnetic phase diagram with complex types of magnetic ordering \cite{Cam2015,Cam1993}. Due to an antiferromagnetic (AFM) coupling at Fe-Gd interfaces and essentially different Curie temperatures of Fe and Gd (for bulk materials, $T_\mathrm{C}^\mathrm{Fe}=1043$~K and $T_\mathrm{C}^\mathrm{Gd}=293$~K) a so-called "compensation point" $T_\mathrm{comp}$ can exist in the system. At $T=T_\mathrm{comp}$ magnetic moments of Fe and Gd layers are equal to each other and the total magnetization of the system vanishes. Below $T_\mathrm{comp}$, the magnetic moment in Gd subsystem exceeds that in Fe subsystem, while above $T_\mathrm{comp}$, opposite situation takes place. As a result, in weak fields applied in the film plane, a collinear magnetic phase is realized with Fe magnetization vector oriented parallel (at $T > T_\mathrm{comp}$) or antiparallel to the field direction (at $T < T_\mathrm{comp}$). As the magnetic field increases and exceeds some critical value, such field-aligned phases become unstable and a transition to canted magnetic state occurs. Moreover, due to a relatively weak exchange stiffness of Gd, the external magnetic field initiates essentially non-uniform distribution of magnetization inside Gd layers (twisted state).

\begin{figure*}
\centering
\includegraphics[width=0.9\textwidth]{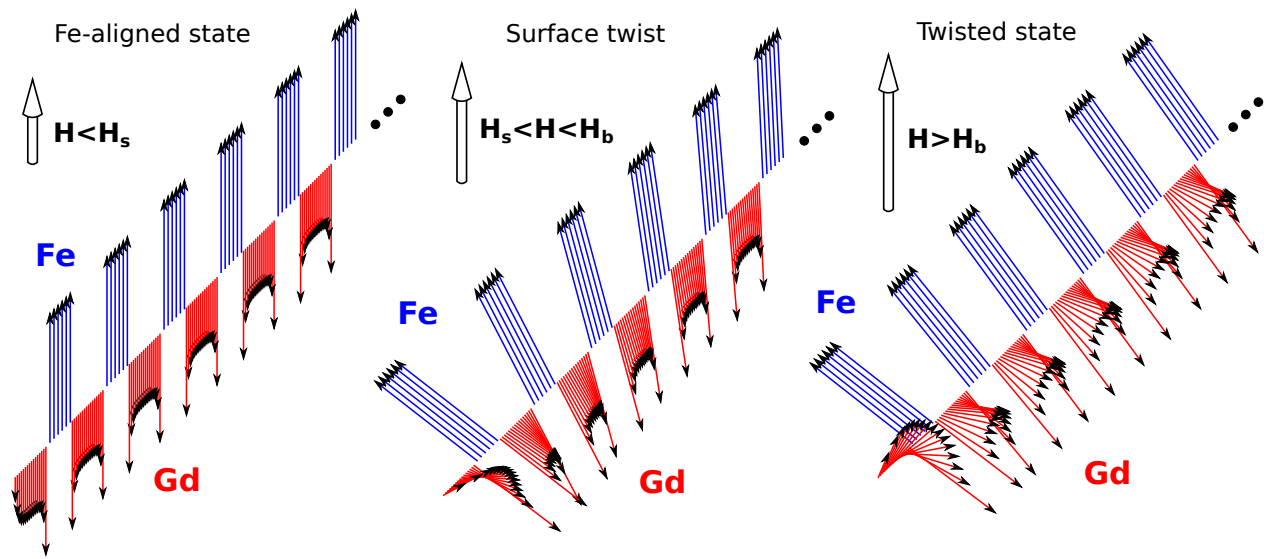}
\caption{Different types of magnetization vector distribution in a Fe/Gd superlattice at $T>T_\mathrm{comp}$ (calculated using the mean-field model \cite{Drov2017}).}
\label{magn_profiles}
\end{figure*}

The above-discussed complex behaviour of the Fe/Gd system was described theoretically by Camley \textsl{et~al.}, using the mean-field approach [3--5], and observed experimentally by different techniques in a number of works [6--9]. At the same time it was predicted theoretically that even more complicated situation takes place when a finite Fe/Gd superlattice is considered. In this case, two types of twisted magnetic states are possible in the system: surface twist and bulk twist \cite{LePage1990}. Starting from the field-aligned state in weak external field, an increase of the field leads first to distortion of the collinear state near the superlattice surface (at $H=H_\mathrm{s}$). At higher fields ($H>H_\mathrm{b}$) the bulk twisted state is realized. Fig.\,\ref{magn_profiles} represents schematically the corresponding magnetization distributions calculated for different field values at $T>T_\mathrm{comp}$ \cite{Drov2017}. It is important to note that the surface twist phase arises at the outermost layer of the superlattice when its magnetization is directed opposite to the applied field. Thus, the surface twist phase arises on Gd-terminated side of the superlattice at $T>T_\mathrm{comp}$ and on Fe-terminated side of the superlattice at $T<T_\mathrm{comp}$.

Direct experimental observation of such surface twisted states comes to difficulties since it requires simultaneous probing bulk and surface magnetic states of the superlattice. A few works were devoted to this problem. Haskel \textsl{et~al.} \cite{Hask2003} demonstrated surface twist effects in a Fe-terminated [Fe/Gd]$_{15}$/Fe multilayer, using grazing-incidence x-ray magnetic circular dichroism. Kravtsov \textsl{et~al.} \cite{Kra2009} used simultaneous refinement of polarized neutron and resonant x-ray magnetic reflectometry data to directly obtain magnetization depth profiles in a [Fe/Gd]$_5$ multilayer. In both cases the complexity of the used methods makes it difficult to perform detailed studies of stability regions of bulk and surface twisted phases as a function of temperature and magnetic field.

Magneto-optical Kerr effect (MOKE) is a relatively simple and sensitive method to obtain direct information about the surface magnetic state of the multilayer. The penetration depth of visible light into metal is about $\sim100$~\AA{} which is comparable with typical thickness of individual layers in the superlattice. Thus, MOKE signal provides information about magnetization in several upper layers of the superlattice. Hahn \textsl{et~al.} \cite{Hahn1995} used MOKE to study surface magnetic states in a [Fe/Gd]$_{15}$ structure. Since the samples were sputtered on non-transparent Si substrates, authors compared MOKE signals from superlattices terminated by Fe and Gd layers. The difference of the MOKE curves for two samples was explained by the surface magnetic twist arising in case of the surface layer magnetization oriented opposite to the field direction.

In our previous work \cite{Drov2017} we studied static magnetization curves of a [Fe/Gd]$_{12}$ multilayer. Comparing the experimental data with mean-field calculations, we found indications of field-dependent phase transitions between field-aligned, surface- and bulk twisted states. However, the static magnetometry provides only the net magnetic moment of the entire multilayer and the surface effects are manifested too weakly. In this work we use MOKE to obtain more precise knowledge on the surface magnetic states in the superlattice. The investigated [Fe/Gd]$_{12}$ multilayer is grown on a transparent glass substrate which allows direct probing magnetic states on both sides of the structure. As a result, we determine the stability regions of bulk and surface twisted states in the superlattice, depending on temperature and magnetic field. The experimental phase diagram is compared with calculations based on the mean-field model \cite{Drov2017}.

Studies of magnetization dynamics in RE/TM systems attract attention due to a recent idea to use such materials for realization of ultrafast magnetic switching, promising for potential applications in magnetic storage devices [14--16]. A number of works were devoted to investigations of ferromagnetic resonance (FMR) in TM/Gd multilayers [17--26]. Room temperature studies [17--20] demonstrated the importance of spin pumping into RE metal to explain a large FMR line width in TM/RE systems. Several groups reported about the effect of line broadening and shift of the absorption peak to lower fields at cooling the system below room temperature. Such behaviour was observed for Co/Gd \cite{Patrin2006,Demirtas2010}, Py/Gd \cite{Khod2017}, Fe/Gd \cite{Drov2017} and Fe/Cr/Gd \cite{Drov2015,Drov2018} multilayers.

In most of the cited works only one "high-temperature" resonance peak was detected. This peak became much weaker or even disappeared as temperature decreased below $T_\mathrm{C}^\mathrm{Gd}$ which effect was explained by non-local damping mechanisms in the system \cite{Drov2017,Khod2017}. In a short letter \cite{Svalov2001}, Svalov \textsl{et~al.} reported about experimental observation of a second absorption peak below $T_\mathrm{C}^\mathrm{Gd}$ in a Co/Gd multilayer. Similar behaviour was observed in our previous works for the Fe/Gd system \cite{Drov2017}. Theoretical simulations showed that the observed absorption peaks corresponded to different types of inhomogeneous resonance modes in the multilayer. In this work we perform more detailed investigation of temperature evolution of the resonance spectra in the Fe/Gd superlattice. In contrast to the work \cite{Drov2017}, here we pay special attention to the transformation of the spectra in the vicinity of $T_\mathrm{C}^\mathrm{Gd}$. In particular, we note that the behaviour of the high-temperature resonance peak is strongly dependent on the pumping frequency. To explain this result and identify the observed resonance modes, the experimental data are compared with model calculations based on Landau-Lifshitz equations describing magnetization dynamics in the system.

\section{Sample and experimental details}

The [Fe(35\,\AA)/Gd(50\,\AA)]$_{12}$ superlattice was prepared on a glass substrate using high vacuum magnetron sputtering technique. Two chromium layers with thickness 50~\AA{} and 30~\AA{} served as buffer and cap layers respectively. X-ray diffraction studies performed in \cite{Drov2017} demonstrated well-defined layered structure of the sample with interfacial root mean square roughness of about 1--2 atomic monolayers.

Magnetic properties of the multilayer were studied using MOKE and FMR techniques in the $4-300$~K temperature range in magnetic fields up to 10~kOe applied in the film plane.

Longitudinal MOKE studies of the surface magnetization were performed on both sides of the film, using a 635~nm semiconductor laser. In our experimental geometry the MOKE signal was proportional to the component of magnetization parallel to the applied field.

FMR measurements were carried out using a conventional field-sweep technique on a laboratory developed transmission type spectrometer at different frequencies in the range $7-36$~GHz.

\begin{figure}[t]
\centering
\includegraphics[width=\columnwidth]{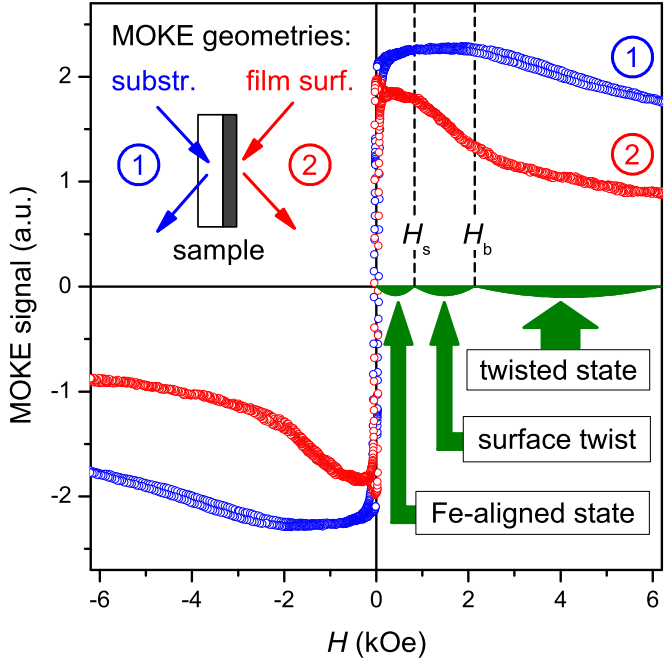}
\caption{MOKE curves measured at 155~K from two sides of the film: 1) from the glass substrate side (Fe-terminated side of the superlattice) and 2) from the film surface (Gd-terminated side of the superlattice). Comparing the curves, different types of magnetic ordering can be identified.}
\label{Kerr}
\end{figure}

\begin{figure}[t]
\centering
\includegraphics[width=\columnwidth]{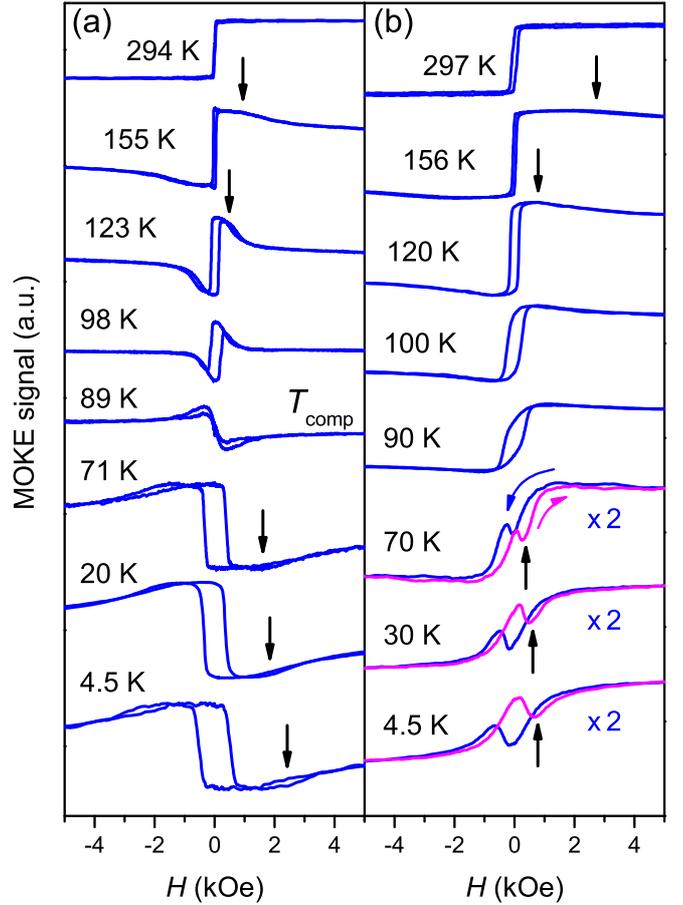}
\caption{MOKE curves obtained at different temperatures on Gd-terminated (a) and Fe-terminated (b) sides of the superlattice. Black arrows show transitions from field-aligned to canted state of the surface magnetization.}
\label{Kerr-T}
\end{figure}

\section{Results and discussion}

\subsection{Magneto-optical Kerr effect}

Static magnetometry of the investigated sample performed in \cite{Drov2017} showed that Gd layers had reduced Curie temperature, $T_\mathrm{C}^\mathrm{Gd}\approx200$~K, comparing with the bulk value 293~K. The system demonstrated the compensation point at $T_\mathrm{comp}\approx90$~K.

Testing MOKE experiments on Fe and Gd thin films showed that both Fe and Gd layers should contribute to the total Kerr effect for the combined Fe/Gd layered system. Under our experimental conditions, the MOKE signal from Gd is comparable with that from Fe (about two times smaller at low temperature) but has opposite sign. Thus, we expect different signs of MOKE for Gd- and Fe-aligned states in the investigated multilayer.

Fig.~\ref{Kerr} shows the experimental MOKE hysteresis loops measured at $T=155$~K from two sides of the superlattice. For both curves, a flat part in the region of weak fields means that the magnetic moment of the outermost layer remains collinear to the external field. Positive sign of the MOKE signal at $H>0$ indicates the Fe-aligned state. At some higher field the MOKE signal decreases, indicating that the magnetization of the outermost layer begins to rotate. Note that on Gd-terminated side this rotation starts in weaker field ($H=H_\mathrm{s}$) than on Fe-terminated side ($H=H_\mathrm{b}$). Thus, we can conclude that in magnetic fields $H_\mathrm{s}<H<H_\mathrm{b}$ the surface twist state is realized on Gd-terminated side of the superlattice. In higher fields $H>H_\mathrm{b}$, a transformation to the bulk twisted phase occurs.

\begin{figure}[t]
\centering
\includegraphics[width=\columnwidth]{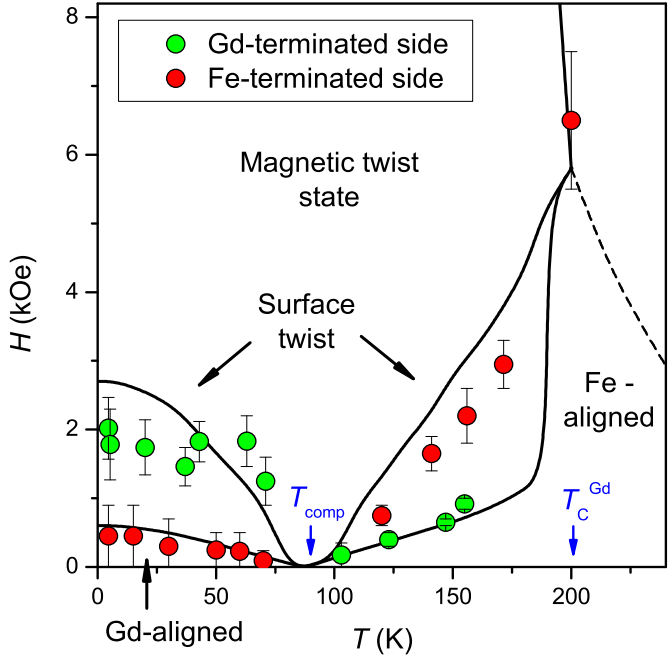}
\caption{Resulting $H-T$ phase diagram of the investigated Fe/Gd superlattice. Points are obtained from MOKE data on two sides of the multilayer. Lines are calculations within the mean-field approach \cite{Drov2017}. The dashed line corresponds to a situation when Gd magnetization vanishes in the middle of Gd layer.}
\label{Phases}
\end{figure}

Similar analysis of the MOKE curves was performed for different temperatures in the range 4--300~K (see Fig.~\ref{Kerr-T}) and the resulting phase diagram of the system was obtained (Fig.~\ref{Phases}). At $T>T_\mathrm{C}^\mathrm{Gd}$ we observe simple rectangular hysteresis loops without any signs of possible phase transitions. At lower temperatures the shape of the MOKE curves changes. The compensation point $T_\mathrm{comp}\approx90$~K can be clearly detected as temperature where an inversion of the hysteresis loop occurs (Fig.~\ref{Kerr-T}), i.e. different orientation of Fe magnetization is realized in weak fields above and below $T_\mathrm{comp}$. It is also clearly seen that at $T>T_\mathrm{comp}$ the rotation of magnetization starts in weaker fields on Gd-terminated side of the superlattice. On the contrary, at $T<T_\mathrm{comp}$ this rotation begins in weaker fields on Fe-terminated side of the multilayer.

Unfortunately, in the region of low temperatures the increasing hysteresis smears the phase transitions and prevents accurate determination of the critical fields. As a result, the experimental error is increasing.

Nevertheless, the observed behaviour is in agreement with the theoretical prediction that the surface twist phase arises on the side of the superlattice when the magnetization of the outermost layer is directed opposite to the applied field. In the phase diagram  Fig.~\ref{Phases}, the experimental stability regions for different phases are compared with the result of mean-field calculations (see \cite{Drov2017} for details). We note a good agreement between the experiment and the model.

\begin{figure*}
\centering
\includegraphics[width=\textwidth]{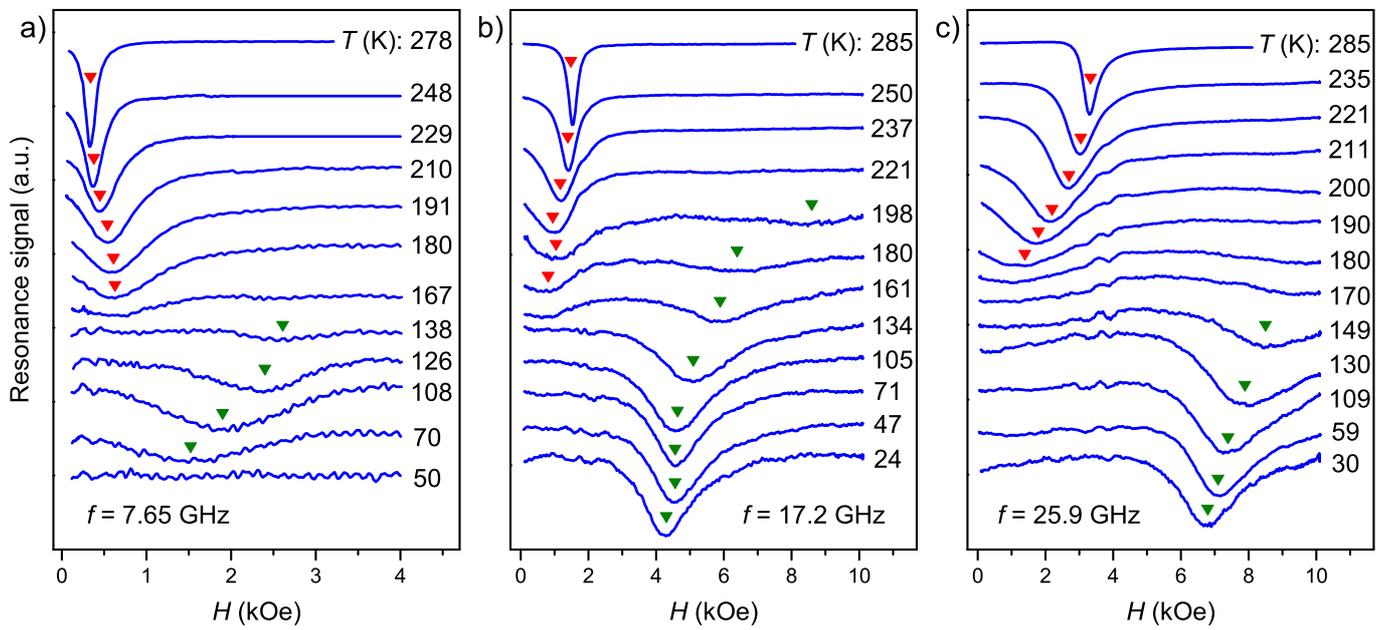}
\caption{Experimental resonance spectra at different temperatures (shown in the plot) obtained at $f=7.65$~GHz (a), $f=17.2$~GHz (b), and $f=25.9$~GHz (c).}
\label{spectra}
\end{figure*}

\begin{figure*}
\centering
\includegraphics[width=\textwidth]{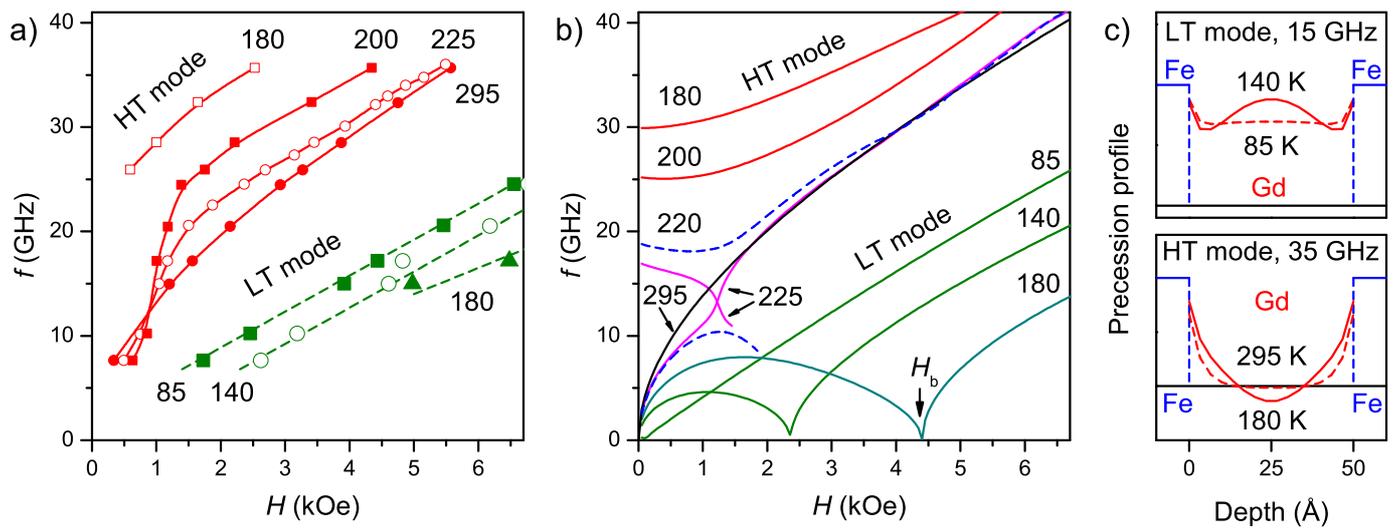}
\caption{Experimental (a) and calculated (b) frequency \textsl{vs} field dependencies at different temperatures (shown in the plots in Kelvins) and examples of calculated depth profiles of magnetization precession in the Gd layer for LT and HT modes (c).}
\label{fvsH}
\end{figure*}

\subsection{Ferromagnetic resonance}

Fig.~\ref{spectra} demonstrates the temperature evolution of experimental resonance spectra at several different frequencies. At room temperature, one relatively narrow ($\Delta H\sim100$~Oe) absorption peak is observed. As temperature decreases, this "high-temperature" (HT) peak broadens and its position changes. We note that the direction of the line shift depends on frequency. At high frequencies ($f\gtrsim12$~GHz) the HT peak shifts towards lower fields (Fig.~\ref{spectra}b,c). The same behaviour was observed earlier for different types of TM/Gd multilayers [21--23]. This effect can be qualitatively described, considering a strongly coupled layered ferrimagnet (see Appendix in the end of this paper), so this behaviour can be considered as "normal".

In our case, however, another situation takes place at low frequencies ($f\lesssim12$~GHz). Here we observe the shift of the HT peak towards higher fields (Fig.~\ref{spectra}a). This result is opposite to the behaviour reported in the previous works [21--23] and clearly contradicts to the simple approximation of strongly coupled FM layers.

At all frequencies under study, the HT peak disappears below $T\approx160$~K. At the same time a second "low-temperature" (LT) peak arises in the region of high fields. As temperature decreases, this peak shifts towards lower fields and becomes more pronounced. At high frequencies it can be clearly detected down to lowest temperature, however, at $f=7.65$~GHz it again disappears below $T\approx60$~K (Fig.~\ref{spectra}).

Fig.~\ref{fvsH}a demonstrates the resulting experimental frequency-\textsl{vs}-field $f(H)$ dependencies at different temperatures. At room temperature the $f(H)$ curve for HT-mode can be qualitatively described by Kittel-like equation for FM film (see Appendix, Eq.~\eqref{A1}). However at lower temperatures the shape of the $f(H)$ curve changes strongly and the simple Kittel's formula clearly becomes inapplicable. This means that the approximation of uniform magnetization precession within the structure is not valid. Taking into account a large exchange stiffness of Fe layers and a strong coupling at Fe-Gd interface, we can suppose that inhomogeneous precession occurs inside the Gd layers. To describe such inhomogeneous resonance modes theoretically we use the approach of the work \cite{Drov2017}.

To model the non-uniform magnetization precession inside Gd layers they are divided into elementary "atomic" sublayers coupled with each other. The static magnetization in each sublayer is calculated using the mean-field model while the dynamics is described by Landau-Lifshitz equations (LLE) with relaxation terms. For relaxation terms, we consider Gilbert damping in Fe and Gd layers as well as diffusion-type damping in Gd (see \cite{Drov2017} for details and model parameters). As a result, we calculate the complex eigenfrequencies of the system $\omega=\omega^\prime+i\omega^{\prime\prime}$. The corresponding eigenvectors represent the depth profiles of magnetization precession in the superlattice. The damping of the calculated resonance modes can be characterized by quality factor (Q-factor) $Q=\omega^\prime/2\omega^{\prime\prime}$. The larger Q-factor is, the more intensive resonance peak is expected. Following our previous work \cite{Drov2017}, we consider only the modes with in-phase precession of Fe layers and perform modelling for one period of the superlattice.

Fig.~\ref{fvsH}b demonstrates the resulting calculated dependencies $f(H)$ at different temperatures (for illustrative purposes, only the modes with $Q>0.5$, are shown). The model predicts the existence of two spectral branches with different types of magnetic precession inside Gd layers (Fig.~\ref{fvsH}c). The HT-mode has a gap in the spectrum at low temperatures and corresponds to strongly non-uniform precession inside Gd layers. The LT-mode is quasi-uniform. Its frequency vanishes at $H=H_\mathrm{b}$, i.e. at phase transition from field-aligned to twisted magnetic state.

In general, the behaviour of calculated curves $f(H)$ repeats qualitatively the experimental dependencies, except the temperature region $\approx200-225$~K (i.e. slightly above $T_\mathrm{C}^\mathrm{Gd}$) and in weak magnetic fields $H\lesssim1.5$~kOe. Above $T=225$~K the model predicts the crossing of two spectral branches. One branch with increasing dependence $f(H)$ corresponds to preferable precession of Fe layers. This branch has large Q-factor and is observed experimentally. The second branch with decreasing dependence $f(H)$ corresponds to preferable precession of inner part of Gd layers. This branch has small Q-factor and is not observed experimentally. Below $T=225$~K the model predicts the repulsion of these two crossing modes. As a consequence, a gap in the spectrum opens. Experimentally, however, such a gap arises only at $T\lesssim180$~K (Fig.~\ref{fvsH}a,b).

Despite this discrepancy, Fig.~\ref{fvsH}b helps to understand different behaviour of the HT peak at frequencies below and above $f\approx12$~GHz, i.e different direction of the line shift at cooling the system below room temperature (Fig.~\ref{spectra}). The critical value 12~GHz corresponds to the frequency where the effect of modes repulsion arises.

\begin{figure}[t]
\centering
\includegraphics[width=\columnwidth]{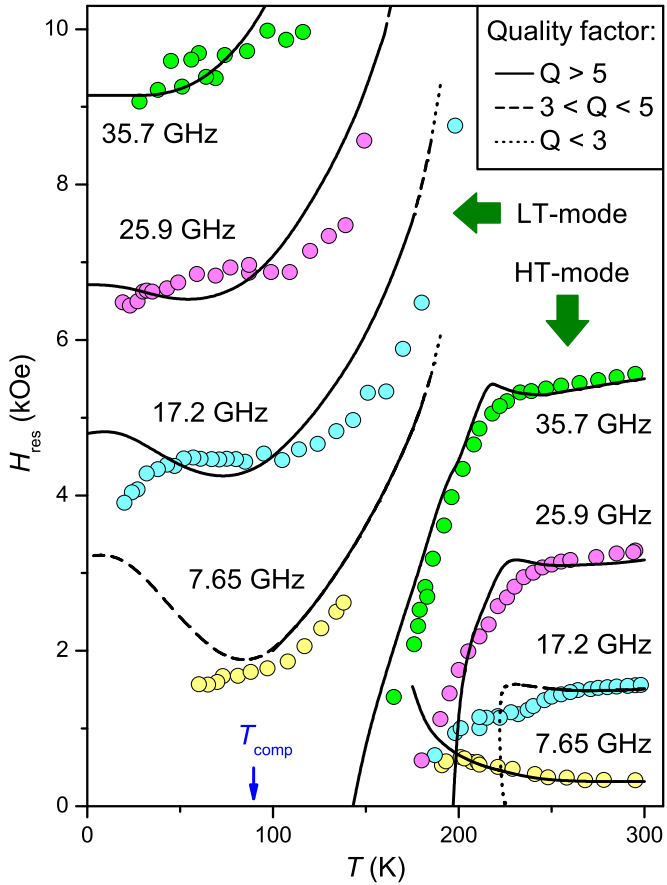}
\caption{Temperature dependencies of the resonance field at different frequencies. Points are experimental data, lines are calculations. Solid, dashed, and dotted lines correspond to different Q-factor of resonance modes.}
\label{Hres-T}
\end{figure}

Fig.~\ref{Hres-T} shows the resulting experimental and calculated temperature dependencies of the resonance fields $H_\mathrm{res}(T)$ at different frequencies. It can be seen that the experimental and theoretical curves demonstrate not only qualitative but also a certain quantitative agreement. The noticeable discrepancy observed for HT-mode at 17.2~GHz below $T\approx230$~K is connected with the above-discussed inadequate description of the mode-repulsion region.

It is interesting to note that at low frequency ($f=7.65$~GHz) the model predicts the existence of minimum in the $H_\mathrm{res}(T)$ dependence for the LT-mode. This minimum is connected with the fact that the LT-mode frequency vanishes at $H=H_\mathrm{b}$ (i.e. $H_\mathrm{res}\rightarrow H_\mathrm{b}$ when $f\rightarrow0$). Since $H_\mathrm{b}$ turns to zero at $T_\mathrm{comp}$, we could expect the minimum of $H_\mathrm{res}(T)$ at this temperature. Experimentally, however, we did not manage to detect the absorption line below $T_\mathrm{comp}$. The reason for this can be the large damping of the corresponding resonance mode. Indeed, our calculations show that the Q-factor of the LT-mode increases below $T_\mathrm{comp}$ at $f=7.65$~GHz (see Fig.~\ref{Hres-T}).

To summarize, we achieved a reasonable agreement between the experiment and model calculations. The model describes many features of the experimental spectra and helps to identify the types of the observed resonance modes. The main discrepancy between the experiment and model arises in the vicinity of $T_\mathrm{C}^\mathrm{Gd}$ where the calculated spectra are very sensitive to magnetic parameters of the system and can be strongly influenced by structural inhomogeneities of the real superlattice.

\section{Conclusion}

In this work we demonstrated the realization of non-colline\-ar magnetic states and inhomogeneous magnetization dynamics in a Fe/Gd artificial layered ferrimagnet. We have shown that both static and dynamic properties of the system are described taking into account essentially non-uniform magnetization distribution inside Gd layers.

Using the magneto-optical Kerr effect, we defined the regions of stability for surface and bulk twisted states of the investigated multilayer. The resulting experimental $H-T$ phase diagram is in a good agreement with calculations based on the mean-field model.

Ferromagnetic resonance spectra obtained in this work reveal a complex temperature evolution with two spectral branches that can not be explained in terms of uniform magnetic precession within the superlattice. The performed theoretical simulations of magnetization dynamics in the system show that the observed resonance modes correspond to different types of inhomogeneous precession inside Gd layers.

In the end we would like to emphasize that the nanostructured ferrimagnets provide possibility to study such complex magnetic phenomena under easily achievable experimental conditions: in magnetic fields up to 1~T and at microwave frequencies. The traditional ferrimagnetic crystals would require magnetic fields and frequencies that are several orders of magnitude larger. In this respect, the artificial structures can be considered as suitable model objects for experimental investigations of non-collinear magnetic phases and inhomogeneous magnetization dynamics in ferrimagnets.

\section*{Acknowledgments}

The work is partially supported by the Russian Foundation for Basic Research (grants No.\,16-02-00061, No.\,18-37-00182), by the Ministry of Education and Science of the Russian Federation (grant No.\,14-Z-50.31.0025), and by the Basic Research Program of the Presidium of Russian Academy of Sciences.

Research in Ekaterinburg was performed in terms of the State assignment of Federal Agency of Scientific Organizations of the Russian Federation (theme ``Spin'' No.~AAAA-A18-188 020290104-2).

\appendix

\section{FMR frequency of a strongly coupled layered ferrimagnet}

Let us consider two FM layers with different magnetic moments $\mu_1 > \mu_2$. We suppose that these layers are strongly AFM coupled (the exchange energy is infinity). In this case, the magnetic field \textbf{H} applied in the film plane aligns $\boldsymbol{\mu}_1$ and $\boldsymbol{\mu}_2$ parallel and antiparallel to the field direction respectively. Considering Zeeman and demagnetizing energy of both layers, the total energy of the system can be written as
$$E=-\mathbf{H}\left(\boldsymbol{\mu}_1+\boldsymbol{\mu}_2\right) + 2\pi\left[ \frac{\left(\boldsymbol{\mu}_1 \cdot \mathbf{z}\right)^2}{V_1} + \frac{\left(\boldsymbol{\mu}_2 \cdot \mathbf{z}\right)^2}{V_2} \right],$$
where \textbf{z} is a unit vector normal to the film plane, $V_1$ and $V_2$ are volumes of layers. Taking into account that $-\boldsymbol{\mu}_2\upuparrows\boldsymbol{\mu}_1$, the energy expression can be rewritten in the form
$$E=-\mathbf{H}\boldsymbol{\mu} + 2\pi\frac{\mu_1^2V/V_1+\mu_2^2V/V_2}{(\mu_1-\mu_2)^2} \cdot \frac{\left(\boldsymbol{\mu} \cdot \mathbf{z}\right)^2}{V},$$
where $\boldsymbol{\mu}=\boldsymbol{\mu}_1+\boldsymbol{\mu}_2$ and $V=V_1+V_2$. Now it has the form of magnetic energy for a single FM film with modified demagnetizing factor. Thus, the FMR frequency of the system is defined by modified Kittel's formula
\begin{equation}
\omega=\gamma_\mathrm{eff} \sqrt{H\left(H+4\pi M_\mathrm{eff} \right)},
\label{A1}
\end{equation}
where
\begin{equation}
4\pi M_\mathrm{eff}=4\pi\frac{\mu_1^2/V_1+\mu_2^2/V_2}{\mu_1-\mu_2},
\label{A2}
\end{equation}
and $\gamma_\mathrm{eff}$ is a net gyromagnetic ratio of two coupled layers \cite{Wan1953}
\begin{equation}
\gamma_\mathrm{eff}=\frac{\mu_1-\mu_2}{\mu_1/\gamma_1-\mu_2/\gamma_2},
\label{A3}
\end{equation}
where $\gamma_1$ and $\gamma_2$ are gyromagnetic ratios of individual layers. If $\gamma_1\approx\gamma_2$, Eqs.~\eqref{A1},~\eqref{A2} predict increasing FMR frequency when $\mu_2$ is increasing. This behaviour is opposite to the case of amorphous or crystal ferrimagnetic film when the effective demagnetizing field is defined by simple expression $4\pi M_\mathrm{eff}=4\pi(M_1-M_2)$, where $M_{1,2}$ are magnetizations of FM sublattices \cite{Wan1953}. In this situation FMR frequency is decreasing with $M_2$ increase.

It is important to note that the approximation \eqref{A1}--\eqref{A3} is valid only when the exchange fields $H_{\mathrm{ex},i}$ acting on layers $i=1,2$ are much stronger than the corresponding demagnetizing fields $H_{\mathrm{ex},i}\gg4\pi M_i$ and the external field is far below the transition to the canted state: $H\ll|H_{\mathrm{ex},1}-H_{\mathrm{ex},2}|$ \cite{Gurevich}.

\end{document}